
\magnification=1200
\baselineskip=20pt
\def\lsim{<\kern-2.5ex\lower0.85ex\hbox{$\sim$}\ }
\def\rsim{>\kern-2.5ex\lower0.85ex\hbox{$\sim$}\ }
\overfullrule=0pt
\def\LAMBDABAR {\hbox{$\lambda$\kern-0.52em\raise+0.45ex\hbox{--}\kern+0.2em}}
\def\ebar {\hbox{E\kern-0.6em\raise0.2ex\hbox{/}\kern+0.1em}}
\ \ \
\vskip 1cm
\centerline{\bf COMMENT ON \lq\lq ATTRACTIVE FORCES BETWEEN}
\centerline{\bf ELECTRONS IN 2 + 1 DIMENSIONAL QED"}
\vskip 1cm
\centerline{by}
\vskip 1cm
\centerline{C. R. Hagen}
\centerline{Department of Physics and Astronomy}
\centerline{University of Rochester}
\centerline{Rochester, NY 14627}
\vfill\eject

It has recently been claimed by Girotti
 \underbar{et} \underbar{al}.$^{1,2,3}$
 that electron--electron bound states are possible
in (QED)$_3$.  They have furthermore used their model to perform extensive
numerical calculations in order to calculate transition temperatures in
high $T_c$ superconductors.  In view of the strong suggestion made by these
authors that the reasonable results obtained are other than mere accident,
it is important to determine whether their proposed model is theoretically
sound.

The point to be made in this Comment consists of the observation that the
binding term of Girotti  \underbar{et} \underbar{al}. is an improper
reduction of the (Aharonov--Bohm) potential $[(\ell +
\alpha_{in}/2)^2 - \ell^2]/r^2$ to simply $\ell \alpha_{in}/r^2$ at large
 $r$.  In fact footnote 4 of ref. 1 makes clear that its authors are aware
of at least one work which includes the additional $\alpha^2_{in}$
repulsive term.  However, they imply that their approach is defensible
because it is \lq\lq based solely on relativistic quantum field theory".

The model of ref. 1 in fact consists of a relativistic perturbation theory
calculation which is then inserted into a Schr\"odinger equation.  Since
the final result thus requires a Galilean limit, it is perfectly sensible
to test the model of ref. 1 by taking its limit $\theta_{in} \rightarrow -
\infty$ with $\theta_{in}/e^2$ finite.  While this eliminates the mass
$\vert \theta_{in}\vert$ particle from the physical spectrum, it has the
advantage of yielding a theory which is exactly soluble as well as fully
relativistic (with respect to the Galilei group of space--time
transformations) and is thus able to provide an unambiguous test of the
proposed model in the indicated limit.

Actually, this problem has been solved some years ago.$^4$  It was shown at
that time that the Hamiltonian of the field theory is
$$H = - {1 \over 2m} \int d^2x \ \psi^+ \left[ \nabla_i + i
g^2 \overline \nabla_i \int {\cal{D}} (x - x^\prime)
 \rho (x^\prime) dx^\prime \right]^2 \psi$$
where $\overline \nabla_i = \epsilon_{ij} \nabla_j, \ g^2 =
 -e^2/\theta_{in}$, and ${\cal{D}} (x)$ is the negative of the inverse
Laplacian.  The operator $\rho$ is the charge density $\psi^+ \psi$
 where $\psi (x)$ satisfies the equal time commutation relation
$$[ \psi (x) , \psi^+ (x^\prime)] =
\delta (x - x^\prime) \quad .$$
For simplicity the spin has been ignored, a feature which can, however,
readily be restored.$^5$

The eigenvalue equation
$$H \vert N>\ = E \vert N>$$
can be solved in the two particle $(N = 2)$ sector.  Then one writes
$$\vert 2 >\ = \int d^2x_1\ d^2x_2\ \psi^+ (x_1) \psi^+ (x_2) f(x_1 , x_2
) \vert 0>$$
and finds in the center--of--mass frame that $f(x_1 , x_2)$ reduces to
$f_\ell (r) e^{i \ell \phi}$ (where $r$ and $\phi$ are the polar coordinates
 of ${\bf  x_1 -  x_2})$.  Significantly, $f_\ell (r)$ satisfies the
equation
$$\left[ {1 \over r} {d \over dr} r {d \over dr} +
 mE - \left( \ell + g^2 / 2 \pi \right)^2 /r^2 \right]
 f_\ell (r) = 0$$
which has a strictly repulsive $1/r^2$ effective potential. With the
identification $g^2 /\pi = \alpha_{in}$ this reproduces the Schr\"odinger
equation of ref. 1 in the $\theta_{in} \rightarrow - \infty$ limit except
for the omission in the latter of the repulsive $(g/2\pi r)^2$ term.  It is
to be emphasized that the present approach fully realizes the ideal of ref.
1 of a purely field theoretical formulation.  This is only partially
realized in ref. 1 where it is necessary to make a crucial junction between
a perturbative field theory calculation and the Schr\"odinger equation.
The result obtained here establishes what would have been anticipated
merely on the grounds of gauge invariance -- namely, that the quadratic
term cannot be neglected.  Thus it is not possible to obtain in a
consistent way the long range attractive potential claimed in refs. 1--3
and the numerical results presented there cannot be expected to be relevant
to the high $T_c$ problem.

\noindent {\bf Acknowledgment}

This work is supported in part by the Department of Energy Grant No.

\noindent DE--FG02--91ER40685.
\vfill\eject
\bigskip
\noindent {\bf References}

\item{1.} H. O. Girotti, M. Gomes, J. L. de Lyra, R. S. Mendes, J. R. S.
Nascimento, and A. J. da Silva, Phys. Rev. Lett {\bf 69}, 2623 (1992).
\item{2.} H. O. Girotti, M. Gomes, and A. J. da Silva, Phys. Lett.
{\bf B274}, 357 (1992).
\item{3.} H. O. Girotti, M. Gomes, J. L. de Lyra, R. S. Mendes, J. R. S.
 Nascimento, and A. J. da Silva, Sept. 1992 preprint.
\item{4.} C. R. Hagen, Phys. Rev. {\bf D31}, 848 (1985).
\item{5.} C. R. Hagen, Int. Jnl.  Mod. Phys. A {\bf 6}, 3119 (1991).

\end